\documentclass[aps,prl,twocolumn,groupedaddress, showpacs]{revtex4-1}
\usepackage[dvips]{graphicx}
\usepackage{bm}
\usepackage{color}
\usepackage{mathtools}

\begin{document}

\title{Effects of Berry Curvature on the Collective Modes of Ultracold Gases}
\author{Hannah M. Price and Nigel R. Cooper}

\affiliation{TCM Group, Cavendish Laboratory, J. J. Thomson Ave., Cambridge CB3 0HE, United Kingdom}

\bigskip

\bigskip

\begin{abstract}

  Topological energy bands have important geometrical properties
  described by the Berry curvature. We show that the Berry
curvature changes the hydrodynamic equations of motion for a
    trapped Bose-Einstein condensate, and causes
significant 
    modifications to the collective mode frequencies. We illustrate
  our results for the case of two-dimensional Rashba spin-orbit
  coupling in a Zeeman field. Using an operator approach, we
  derive the effects of Berry curvature on the dipole mode in
  very general settings. We show that the sizes of these effects can
  be large and readily detected in experiment. Collective modes therefore provide a
    sensitive way to measure geometrical properties of
    energy bands.

\end{abstract}

\pacs{03.65.Vf, 67.85.-d, 03.75.Kk}  

\maketitle

Nontrivial topological energy bands exhibit many fascinating physical phenomena. For instance, topological invariants underlie both the quantum Hall effect\cite{thouless} and topological insulators\cite{hasankane,qizhang}. There is currently great interest in exploring such physics in ultracold gases\cite{bdn}. Recent experiments have explored optical lattices with novel geometrical and topological features\cite{blochstaggered,esslinger,strucksengstock,blochuniform,ketterleuniform}, and there is much ongoing activity to extend to other situations.

Less widely appreciated is the fact that the energy bands of these new forms of optical lattice also have important {\it geometrical} properties. In particular, the Berry curvature (defined below) is a local, geometrical property of the energy eigenstates. When integrated over the Brillouin zone (BZ) of a two-dimensional (2D) band, it gives the Chern number, the topological invariant of the quantum Hall effect\cite{thouless,barrysimon}. The Berry curvature has many physical consequences in 2D and 3D systems, such as in the anomalous quantum Hall effect\cite{jungwirth,onoda,haldanefs,di}. In ultracold gas experiments, the local Berry curvature could be measured directly, for instance, in the semiclassical dynamics of a wave packet\cite{price} or in time-of-flight measurements\cite{pachos}.

In this Letter, we show that the Berry curvature crucially affects the collective modes of an ultracold gas. 
Thus, this geometrical quantity must be added in the general theory of collective modes to describe the new forms of optical lattice currently being explored.
Collective modes are powerful tools for exploring the properties of ultracold gases\cite{pethick}. The high precision with which oscillation frequencies can be measured~\cite{altmeyer} affords high sensitivity to underlying physical properties, such  as the equation of state\cite{griffin,bruun} and the BEC-BCS crossover\cite{bartenstein,heiselberg,stringarifermions}.
 Recently, collective modes have been used to measure the superfluid Hall effect in a weak artificial magnetic field\cite{leblanc}. 
Here, we show that, in the general case of an atomic gas in a band structure with geometrical features, there are important modifications of the collective modes which are entirely controlled by the Berry curvature. This includes systems threaded with many flux quanta, as well as optical lattices with uniform \cite{JakschZoller} or nonuniform flux \cite{blochstaggered}.
 We demonstrate how the Berry curvature shifts oscillation frequencies and splits otherwise degenerate modes. We illustrate this for the example of 2D Rashba spin-orbit coupling. We then derive the effects of Berry curvature for a general multiband Hamiltonian. Our results show that Berry curvature can have large effects on collective mode frequencies of  trapped BECs, and that measurements of these frequencies can be used to determine the geometrical properties of the energy band in which the condensate is formed.

The starting point for all our studies is a Bose-Einstein condensate
(BEC) formed in a minimum of some single-particle energy
  dispersion, $E({\bm p})$. This dispersion could be the lowest band
  of an optical lattice, with ${\bm p}$ the crystal momentum, or of a spin-orbit coupling Hamiltonian [like
  Eq. (\ref{eq:H})]. We shall study the collective oscillations of the
  BEC confined in a (harmonic) trap.  We assume that the band gap is much larger than any other energy
scale, and that the spread in momentum of the condensate wave function
(set by the inverse cloud radius or healing length) is sufficiently
small\footnote{In the Thomas-Fermi approximation, we require interactions to be sufficiently weak that $\xi \gg a$, where $\xi = ({\hbar^2/ 2 M \rho_0 g})^{1/2}$ is the healing length and $a$ is the optical lattice spacing. For the Rashba model discussed in the text, the effective mass approximation breaks down for momenta $p \simeq \Delta / \lambda$, and so we require that the healing length is much greater than $\hbar \lambda/\Delta$.}, such that the BEC is well described by
single-particle states close to this single minimum.  The energy
dispersion is then characterized by
the effective mass, $M^*_{\alpha\beta} = \hbar^2 /\left(\partial^2
  E/\partial p_{\alpha} \partial p_{\beta} \right) $, where $\alpha,
\beta$ run over the spatial coordinates. This will shift the
collective mode frequencies\cite{kramer}. Furthermore, the
eigenstates are characterized by the Berry curvature,
\begin{eqnarray} 
{\bm \Omega}_n({\bm p})& \equiv & i \frac{\partial}{\partial {\bm p}} \times  \langle n{\bm p}| \frac{\partial}{\partial {\bm p}} |n{\bm p}\rangle , \label{eq:omega}
\end{eqnarray}  
where $|n{\bm p}\rangle$ is the energy eigenstate in band $n$ at ${\bm p}$ (i.e., the periodic Bloch function for an optical lattice)\cite{di}. Thus, the energy minimum must also be characterized by the value of the Berry curvature at that point.
To simplify presentation, we assume that the effective mass is isotropic, $M^*_{\alpha\beta} \equiv M^* \delta_{\alpha\beta}$, and choose axes such that the local Berry curvature is $\Omega \hat{\bm z}$,  but all results can be readily extended to anisotropic cases.

To determine the effects of Berry curvature on the collective modes, we
derive
the hydrodynamic equations at zero temperature for a weakly interacting BEC. By including the so-called ``anomalous contribution'' to the velocity~\cite{di} we find 
\begin{eqnarray} 
&&\dot{ \rho} + {\bm \nabla} \cdot (\rho {\bm v})=0 , \qquad \dot{{\bm v}}= \frac{{\bm F}}{M^{*}}-\left(\frac{\dot{{\bm F}}}{\hbar} \times \hat{{\bm z}}\right) \Omega  \label{eq:mot}
\end{eqnarray}
where $\rho$ is the density, ${\bm v}$ is the velocity, and ${\bm F}$ is the local force per particle. We are interested in small deviations from equilibrium, $\rho = \rho_0+\delta \rho$. We assume that the particle number $N$ is large so that quantum pressure is negligible and the Thomas-Fermi approximation is valid: $\rho_0= [ \mu - V({\bm r})]/ g $, where $V({\bm r})$ is the trapping potential and $\rho g$ is the interaction energy. Then ${\bm F} =-g {\bm \nabla} \delta \rho$. Linearizing Eq. (\ref{eq:mot}) with respect to $\delta \rho$, we find
\begin{eqnarray}
\delta\ddot{ \rho} =- \frac{{\bm \nabla} V \cdot  {\bm \nabla} \delta \rho}{M^*} +\frac{ \rho_0 g {\bm \nabla}^2 \delta \rho}{M^{*}}+\frac{{\bm \nabla} V \cdot ({\bm \nabla} \delta \dot{\rho} \times\hat{{\bm z}}) \Omega }{\hbar} . \label{eq:master}
\end{eqnarray}
For a uniform gas, with no trap potential $V({\bm r})=0$, the
  collective oscillations are sound modes, with frequencies $\omega =
  \sqrt{\rho_0 g / M^*} |{\bm k}|$ that are unaffected by Berry
curvature.

For a harmonic trap, $V({\bm r})=\frac{1}{2} \kappa |{\bm r}|^2$, the modes have the form $\delta \rho = D(r) Y_{lm} (\theta, \varphi)e^{ - i \omega t }$, where $Y_{lm} (\theta, \varphi)$ is a spherical harmonic\cite{stringarioriginal,stringarireview}. There are three quantum numbers: $l$, the total angular momentum, $m$, the projection of angular momentum on the polar axis, and $n_r$, the number of radial nodes. We solve Eq. (\ref{eq:master}) to find 
\begin{eqnarray} 
\omega &=&- \frac{  m \kappa  \Omega}{2 \hbar} \nonumber \\ &&+ \frac{1}{2} \sqrt{ \left(\frac{  m \kappa \Omega}{\hbar}\right)^2 + \frac{4 \kappa}{M^*} ( l + 3n_r + 2n_r l+2n_r^2)} . \label{eq:freq}
\end{eqnarray}
The corresponding eigenstates have $D(r) \propto  r^l F(- n_r , l + n_r + 3/2; l+ 3/2, r^2/R^2)$,
where 
 $F$ is the hypergeometric function and $R=\sqrt{2\mu/\kappa}$ the radius of the cloud. When $\Omega=0$, we recover the expected mode energies \cite{stringarioriginal,stringarireview}. Nonzero Berry curvature affects the frequencies of only those modes with $m\neq 0$, breaking the $(2m+1)$ degeneracy.

 An important class of modes are the surface waves which have $n_r=0$
 and $\delta \rho \propto r^{l-1} Y_{lm} (\theta, \varphi)e^{ -
     i \omega t } \frac{\partial \rho}{\partial r}$. These include
 the dipole modes ($l=1$) and quadrupole modes ($l=2$). We find that,
 as in the case without Berry curvature \cite{griffin,bruun}, the
 mode frequencies are independent of the equation of state. We obtain
 this result by extending the hydrodynamic approach to a general
 polytropic equation of state: $P \propto \rho^{\gamma +1} $, where
 $P$ is the pressure and ${\bm F} = - \frac{1}{\rho}{\bm \nabla}P -
 {\bm \nabla} V \nonumber$. For the weakly interacting Bose
 condensate, $P = \frac{1} {2} g \rho^2 $ and $\gamma =1$. The
 exponent $\gamma=2/3$ describes a dilute Fermi gas\footnote{Our results can be extended to Fermi systems,
   provided the atoms only occupy states near the minimum of the
   energy dispersion, and so are described by the same effective mass and
   Berry curvature.}, an ideal normal
 Bose gas under adiabatic conditions and Bose and Fermi gases in the
 strongly interacting (unitarity) limit \cite{griffin,bruun,heiselberg}.

\begin{figure}

  \centering
\resizebox{0.45\textwidth}{!}{\includegraphics*{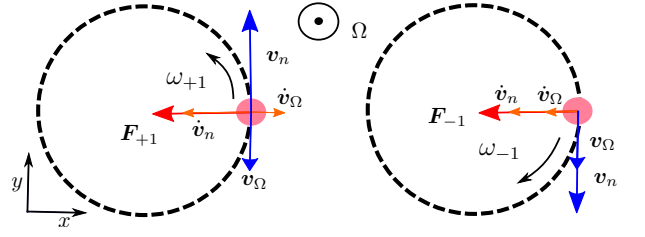}}  
\caption{The Berry curvature splits degenerate modes, as shown here for the 2D dipole oscillations in a weakly interacting Bose gas: $\delta \rho_{\pm 1} = (x \pm i y ) e^{-i \omega t}$ (i.e. $n_r=0$, $l=1$, $m=\pm 1$). An element of fluid (pink circle) feels a restoring force, ${\bm F}_{\pm 1} =-g {\bm \nabla} \delta \rho_{\pm 1} = - g \left( 1 , \pm i \right)e^{-i \omega t}$, around the dashed curve. Without Berry curvature, the acceleration, $\dot{{\bm v}}_n$, and velocity, ${\bm v}_n$, have the same magnitude for both modes and $\omega_{+1}=\omega_{-1}$. The Berry curvature couples to the time-dependent force, giving an additional acceleration $\dot{{\bm v}}_{\Omega} $ and velocity ${\bm v}_{\Omega} \propto  ({\bm F} \times \hat{{\bm z}}) \Omega$. This is analogous to the ``guiding center" velocity of a particle in electromagnetic fields [c.f. ${\bm v} \propto  ({\bm E} \times {\bm B}) $]. The resultant frequencies are split; $\omega_{+1}$ is lowered and $\omega_{-1}$ is raised.} \label{fig:dipole} 
\end{figure}

Surface modes are also valid solutions for an anisotropic trap: $V(x,y,z) = \frac{1}{2} \kappa^2 (x^2+y^2) + \frac{1}{2}\kappa_z^2  z^2$. The anisotropy lifts the degeneracy between modes with different values of $|m|$. For example, without Berry curvature, the dipole modes for the weakly interacting BEC are $\delta \rho \propto (x \pm i y) \propto r Y_{1\pm 1} (\theta, \varphi)$ at frequency $\omega= \sqrt{\kappa/M^*}$, and $\delta \rho \propto z \propto r Y_{1 0} (\theta, \varphi)$ at frequency $\omega=\sqrt{\kappa_z/M^*}$ \cite{stringarireview}. In what follows, we shall refer to $\delta \rho \propto (x \pm i y)$ as the (quasi-) 2D dipole modes. Figure \ref{fig:dipole} shows schematically how the Berry curvature splits these modes. In the absence of Berry curvature, the quadrupole modes with $m=\pm 2$ or $m= \pm1$ can also be linearly combined to give the scissors modes \cite{scissors}(with
 $\delta \rho \propto xy, yz, xz$). The existence of the scissors modes relies on the degeneracy between $\pm |m|$. Here these mode frequencies are split, and so the modes must retain their angular symmetry. 

Another important low-lying mode is the breathing mode ($n_r=1$ and $l=m=0$). Without Berry curvature, the velocity field is purely radial (${\bm v} \propto {\bm r}$), and the density oscillation resembles a ``breathing" of the cloud \cite{stringarireview}. The mode frequency now depends on the equation of state: $\omega = \sqrt{ (3 \gamma +2)\kappa/M^* }$\cite{griffin,bruun}. Our results show that this mode frequency is unchanged for $\Omega \neq 0$, since it has no angular momentum along ${\bm \hat{z}}$. However, it is interesting to note that the mode velocity field is changed, gaining an extra rotational (divergence-free) component $\propto {\bm r} \times \hat{{\bm z}}$.

We illustrate the effects of the Berry curvature on collective modes for a simple model of Rashba spin-orbit coupling. Recent experiments have studied spin-orbit coupling in 1D \cite{Lin,zhangdipole} and there are proposals for extensions to two dimensions \cite{dalibard,andersonrashba,xuueda}. We consider a 2D interacting spin-1/2 gas described by the Hamiltonian
\begin{eqnarray}
&&\hat{H} = \sum_i \hat{h}_0(i) +
\frac{1}{2}{g_{\rm 2D}}  \sum_{i \neq j}  \delta(x_i-x_j) \delta(y_i-y_j)
\nonumber \\ &&
\hat{h}_0= \frac{{\bm p}^2}{2M} + \lambda (p_x\hat{\sigma}_y - p_y \hat{\sigma}_x) - \Delta \hat{\sigma}_z + V(x,y) . \label{eq:H}
\end{eqnarray}
where $g_{\rm 2D}$ is the effective contact interaction in two dimensions, $i=1,...,N$ is the particle index, and $\hat{\sigma}_{x,y,z}$ are the Pauli matrices. We assume that the interaction strength is independent of spin, which is a good approximation for $^{87}$Rb. The single-particle Hamiltonian, $\hat{h}_0$, is characterized by a Rashba spin-orbit coupling, $\lambda$, and a Zeeman field, $\Delta$. The effects of spin-orbit coupling on the collective modes have previously been studied for one dimension \cite{zhangdipole,martone,stringari} (where there can be no Berry curvature) and for the 2D dipole mode in a thermal gas using a Boltzmann approach \cite{duine}.   

Without a trap, the single-particle energy spectrum is $\varepsilon_{\pm} = \frac{p^2}{2M}\pm \sqrt{\lambda^2 p^2 + \Delta^2}$. When $\zeta \equiv \frac{\lambda^2 M}{\Delta} < 1$, there is a single minimum in the lower band at $p=0$. This minimum has effective mass $M^* = \frac{M}{(1- \zeta)}$, and  Berry curvature $\Omega = \frac{\lambda^2 \hbar^2}{2 \Delta^2}$  \cite{di}.
We consider the collective oscillations of a BEC formed in this single minimum.
In addition to $\zeta$, the mean-field theory for the Hamiltonian [Eq. \ref{eq:H}] is characterized by two other dimensionless parameters: $\chi \equiv \frac{\hbar \omega_0}{\Delta}$ (where $\omega_0 = \sqrt{\kappa/M}$), which compares the trap and the band gap, and $G \equiv \frac{ N g_{\rm 2D} M }{\hbar^2} $, which is a measure of the interaction strength. We assume that $G \gg 1$ to justify the Thomas-Fermi approximation which improves with increasing particle number, $N$. We also take $\chi \lesssim 1$, to avoid mixing with higher bands. 

\begin{figure} [htdp]
\centering
\resizebox{0.45\textwidth}{!}{\includegraphics*{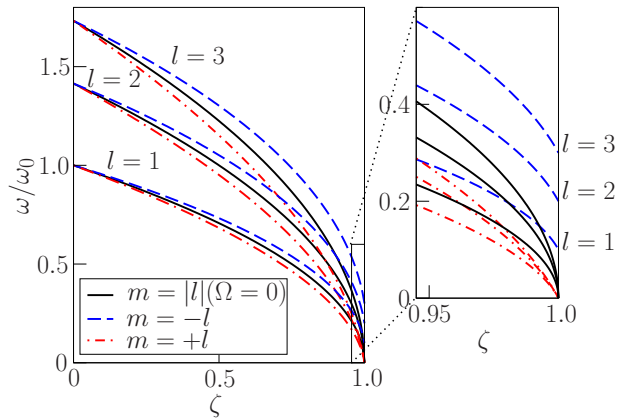}}  
\caption{The three lowest-energy sets of 2D surface modes ($n_r=0$, $l= |m|= 1,2,3$) for the Rashba Hamiltonian for a weakly interacting BEC. The modes are shown in the single-minimum regime, $0<\zeta<1$, for intermediate trapping, $\chi=0.2$ (the $\Omega =0$ results are for weak trapping $\chi \ll 1$). The Berry curvature breaks the degeneracy and splits the modes.} \label{fig:rashba}
\end{figure}

The three lowest sets of 2D surface mode frequencies [$n_r=0$, $l=|m|=1,2,3$ in Eq. (\ref{eq:freq})] are shown in Fig. \ref{fig:rashba} for intermediate trapping, $\chi=0.2$, where the splitting due to Berry curvature is significant. Without Berry curvature (in the limit $\chi \ll 1$), the mode frequencies are $\omega/\omega_0=\sqrt{(M/M^*)l}$ for both $m=\pm l$. As $\zeta \rightarrow 1$, the effective mass, $M^* = \frac{M}{(1- \zeta)}$, diverges and $\omega/\omega_0 \rightarrow 0$. This is the transition from the single minimum to the ring of degenerate minima at nonzero momenta in the energy spectrum. When Berry curvature is present, the splitting between surface modes with $m= \pm l$ is $\frac{\delta \omega}{\omega_0}= \frac{1}{2} \zeta \chi l$  [Eq. (\ref{eq:freq})]. It is also interesting to note that, while the mode frequencies without Berry curvature all go to zero at $\zeta=1$ (where $M^* \rightarrow \infty$), for nonzero Berry curvature there remain modes at nonzero frequency\footnote{For clarity, we have not shown higher-lying modes in Fig. \ref{fig:rashba}, although some of these go soft as the effective mass diverges.}. 

For the particular case of the dipole mode, we now show how the
effects of Berry curvature arise naturally for a very general multiband
system. We assume that the minimum of the lowest band is
at a high symmetry point (e.g. the Brillouin zone center, ${\bm
    p}={\bm 0}$). We choose axes such that the local Berry curvature is $\Omega \hat{{\bm z}}$, and discuss motion in the $xy$
plane (motion along $z$ decouples). The single-particle Hamiltonian is
$H= - \frac{\hbar^2}{2M}{\bm \nabla}^2 + V({\bm r}) + U({\bm r})$, where
$U({\bm r})$ is the periodic lattice potential. Following the usual effective mass theory\cite{luttinger}, we expand the wave function in terms of functions 
$e^{i{\bm k}\cdot{\bm r}}|n\rangle$,
  where $|n\rangle \equiv |n {\bm 0}\rangle$ is the Bloch
    function of the $n$th band at ${\bm p}={\bm 0}$.
We find
\begin{eqnarray}
\hat{H} &=& \sum_n\left(E_n +\frac{\hbar^2|{\bm \hat{{\bm k}}}|^2}{2M} +\frac{1}{2}M \omega_0^2 |{\bm \hat{{\bm x}}}|^2 \right) |n\rangle\langle n|
\nonumber \\
&& + \sum_{\alpha, n, n'} \frac{\hbar\hat{k}^\alpha}{M} \pi_{n,n'}^\alpha 
|n\rangle \langle n'|  
\end{eqnarray}
where $E_n$ is the $n$th band energy at the zone center, and $\alpha$
runs over spatial dimensions.  The operator $\hbar\hat{{\bm k}}$ is the
  crystal momentum (with eigenvalues $\hbar{\bm k}$ for the states
  $e^{i{\bm k}\cdot{\bm r}}|n\rangle$) and $\hat{{\bm x}}$ is the
conjugate crystal position. The last term in $\hat{H}$ is of
  the familiar ``${\bm k} \cdot {\bm p}$" form~\cite{luttinger}, with
  $\pi ^\alpha_{n,n'} \equiv \langle n | \hat{ p}^\alpha | n' \rangle$
  the interband matrix elements of $\hat{p}$
[N.B. $\pi^{\alpha}_{n,n}=M (\partial E_n / \partial p^{\alpha} )= 0$
at a band minimum].

For particles in a {\it quadratic} band minimum, the dipole mode is a center-of-mass oscillation. It is therefore unaffected by interactions (which depend only on interparticle separations). As we now show, the dipole mode remains a center-of-mass oscillation for nonzero Berry curvature. Hence, interactions can be neglected provided they do not excite particles to high-energy states which lie beyond the effective mass approximation, which we now assume. The Heisenberg equations of motion are then
\begin{eqnarray}
\hat{\dot{X}}^\alpha & = & \frac{1}{M} \hat{P}^\alpha  + \frac{1}{M}  \sum_{n, n'} \hat{\Pi}^\alpha_{n n'}  , \qquad
\hat{\dot{P}}^\alpha  =  - M \omega_0^2 \hat{X}^\alpha \nonumber \\
\hat{\dot{\Pi}}^\alpha_{nn'} & = & \frac{i}{\hbar} (E_n - E_{n'} ) \hat{\Pi}^\alpha_{n n'}  \nonumber \\ &&+ 
\frac{i}{\hbar} \frac{\hat{P}^\beta}{M} \sum_j  \pi_{nn'}^\alpha \left( \pi_{jn}^\beta |j\rangle \langle n'|
- \pi_{n'j}^\beta |n\rangle \langle j|\right) 
\end{eqnarray}
for the crystal position, $\hat{X}^\alpha \equiv  \hat{x}^\alpha \sum_n |n\rangle \langle n|$, crystal momentum, $\hat{P}^\alpha  \equiv  \hbar\hat{k}^\alpha \sum_n |n\rangle \langle n|$, and Bloch momentum, $\hat{\Pi}^\alpha_{nn'}  \equiv  \pi^{\alpha}_{nn'} |n\rangle \langle n'| $ (all defined at $t=0$). 

These equations describe coupling of motion of the center-of-mass to interband transitions.
Assuming that all interband transition energies $E_n-E_0$ are large compared to $\hbar \omega$ (so all atoms are in the lowest band, $n=0$),  we can approximate the last line, replacing $\hat{P}^\beta\hat{O}$ with $\hat{P}^\beta \langle 0| \hat{O}  |0\rangle $. This line becomes $\propto \hat{P}^\beta \left(  \pi^\beta_{0n} \pi_{n0}^\alpha \delta_{n'0}- \pi_{n'0}^\beta \pi_{0n'}^\alpha \delta_{n0}\right)$, such that only $\hat{\Pi}^\alpha_{n0}$
and $\hat{\Pi}^\alpha_{0n'}$ couple to $\hat{P}$. Taking the operators to vary harmonically with $e^{-i \omega t}$, and eliminating  $\hat{\Pi}^\alpha_{nn'}$, we find 
\begin{eqnarray}
- i&& \omega \hat{{P}}^\alpha  =  - M \omega_0^2 \hat{X}^\alpha \nonumber\\
-i&&\omega \hat{{X}}^\alpha  \simeq  \frac{1}{M} \hat{P}^\alpha  \nonumber \\ &&
- \frac{\hat{P}^\beta }{M^2} \sum_{n>0} \left[
 \frac{\pi^\beta_{0n}\pi^\alpha_{n0} }{
\left[\hbar\omega + (E_n-E_0)\right]}
-
 \frac{\pi^\alpha_{0n}\pi^\beta_{n0}}{
\left[\hbar\omega - (E_n-E_0)\right]}\right]  . \label{eq:intermediate}
\end{eqnarray}
We expand this to first order in $\hbar \omega /  (E_n-E_0)$ and find
\begin{eqnarray}
-i\omega \hat{X}&\simeq & \left( \frac{1}{M^*}\right)_{xx}\hat{P}^x  +\left( \frac{1}{M^*} \right)_{xy}\hat{P}^y  +  \frac{ i \omega\Omega}{\hbar}\hat{P}^y \nonumber \\
-i \omega \hat{Y} & \simeq  & \left( \frac{1}{M^*}\right)_{yy}\hat{P}^y  +\left( \frac{1}{M^*} \right)_{yx}\hat{P}^x   - \frac{ i \omega\Omega}{\hbar}\hat{P}^x \label{eq:motion}
\end{eqnarray}
where we have introduced the effective mass \cite{callaway} and Berry curvature \cite{di} for the lowest band at the Brillouin zone center
\begin{eqnarray}
\left(\frac{1}{M^*}\right)_{\alpha\beta} &\equiv& \frac{1}{M}\delta_{\alpha\beta} -\frac{1}{M^2}
\sum_{n>0} 
 \frac{\pi^\beta_{0n}\pi^\alpha_{n0}+\pi^\alpha_{0n}\pi^\beta_{n0}}{(E_n-E_0)} \nonumber \\ 
\Omega &\equiv &\frac{i\hbar^2}{M^2}\sum_{n>0} 
 \frac{\pi^x_{0n}\pi^y_{n0}-\pi^y_{0n}\pi^x_{n0}}{(E_n-E_0)^2} .   \label{eq:define}
 \end{eqnarray}
This expression for $\Omega$ can be derived from Eq. (\ref{eq:omega}) using the relation $\langle n | \frac{\partial H}{\partial {\bm p} } | n'\rangle =  \langle \frac{\partial n }{\partial {\bm p}} |  n'\rangle (E_n -E_{n'})$ \cite{di}. 
From Eq. (\ref{eq:motion}), we calculate the dipole frequencies of $\hat{D}_{\pm 1} = \hat{X} \mp i \hat{Y}$ (corresponding to the modes $\delta \rho_{\pm1}$). For an isotropic effective mass, the dipole mode frequencies are given by Eq. (\ref{eq:freq}) (with $n_r=0$, $l=1$ and $m=\pm1$), confirming the hydrodynamic result. Moreover, since all atoms oscillate in the same way, the dipole mode remains a center-of-mass oscillation, so this result is independent of the regime or equation of state. Our derivation shows that the Berry curvature appears as the next-order correction after the effective mass\footnote{Away from high symmetry points, there can also be cubic corrections to the effective mass which will scale on the same order as the Berry curvature [$\propto 1/(E_n-E_0)^2$].}. 

Finally, we discuss how the observation of collective modes frequencies may be used to experimentally characterize the geometrical properties of energy bands. The Berry curvature can be directly measured in the frequency splitting between those modes which are degenerate for $\Omega=0$. For example, the splitting between surface modes with $m= \pm l$ is given by $\frac{\delta \omega}{\omega_0}= l {\Omega}/{a_0^2}$ (Eq. \ref{eq:freq}), where $a_0 = \sqrt{{\hbar}/{M \omega_0}}$ is the harmonic oscillator length. This splitting will manifest as a {\it precession} of distortions formed from $+m$ and $-m$ at frequency $\delta \omega$.

The sizes of the effects we predict can be very large. For example, 
in the 2D Rashba model studied above, the energy splitting of the modes can be written $\frac{\delta \omega}{\omega_0}= \frac{1}{2} \zeta \chi l$. As previously discussed, it is necessary that $\chi \lesssim 1$ and $0<\zeta<1$; this sets an upper limit of $\frac{\delta \omega}{\omega_0}\lesssim 50 \% \times l$. Thus, the energy splitting can be as large as the trap frequency itself.

For optical lattices that cause the atoms to experience an average magnetic field, the effects of  Berry curvature become stronger at {\it smaller} flux densities. To see this, consider
the Harper-Hofstadter model of a tight-binding lattice with a flux $n_{\phi}=1/q$ (where $q$ is an integer) per plaquette of dimensions $a\times a$ \cite{hofstadter}. The magnetic unit cell contains one flux quantum, so its area is $A = a^2 /n_\phi$. The corresponding magnetic BZ  has an area $A_{\rm BZ}= (2 \pi)^2 n_\phi / a^2$. The average Berry curvature $\bar{\Omega}$ scales as $ \propto 1/ n_\phi$ (because $\bar{\Omega}  A_{\rm BZ} = 2 \pi C$, where $C$ is the Chern number, which is $C=1$ in this case). Hence the effects of Berry curvature increase for small $n_\phi$.
That said, these effects can be large even when the flux per unit cell is of order 1. For optical flux lattices \cite{nigel,nigelnew}, and taking $^{87}$Rb atoms condensed in the minimum of the $F=1$ two-photon coupling scheme [with parameters of Fig. 4(a) in Ref.~\onlinecite{nigelnew}], the splitting between two surface modes with $m=\pm l$ is $\delta \omega/ \omega_0 \approx 3.4\% \times l $ for $\omega_0/2\pi=150$Hz [Eq. (\ref{eq:freq})]. For the quadrupole modes (with $m=\pm2$), $\delta \omega/2\pi \approx 10$Hz. This is larger than measured damping rates and well within typical experimental measurement precision\cite{dampingkurn}.

In conclusion, we have shown that Berry curvature has important effects on the collective modes of ultracold gases.
We derived the general hydrodynamic theory for collective modes including Berry curvature, and illustrated its effects for situations of current experimental interest\cite{blochuniform,ketterleuniform}.
The Berry curvature can lead to large splittings of mode frequencies, which should be readily detectable with current experimental capabilities. Their observation would allow a characterization of the geometrical properties of BECs in topological energy bands.\\

\acknowledgments{We are grateful to Stefan Baur for helpful conversations.
This work was supported by the EPSRC.}

\end{document}